\documentclass[sigconf]{acmart}
\acmConference[ESEC/FSE 2021]{The 29th ACM Joint European Software Engineering Conference and Symposium on the Foundations of Software Engineering}{23 - 27 August, 2021}{Athens, Greece}

\usepackage{algorithm} 
\usepackage{algpseudocode} 
\usepackage{stfloats}
\usepackage{fnpos}
\usepackage[bottom]{footmisc}
\usepackage{tabularx}
\usepackage{pifont}
\usepackage{tabularx}
\usepackage{amsmath}
\usepackage{listings}
\usepackage{multirow}
\usepackage{makecell}
\usepackage{xcolor,colortbl}     
\usepackage{lipsum}     
\usepackage[framemethod=TikZ]{mdframed}
\usepackage{subcaption}
\usepackage{enumitem}
\usepackage{tikz}

\renewcommand\footnotetextcopyrightpermission[1]{} 

\newcommand*\emptycirc[1][1ex]{\tikz\draw[thick] (0,0) circle (#1);} 

\newcommand*\fullcirc[1][1ex]{\tikz\fill (0,0) circle (#1);} 

\def\BibTeX{{\rm B\kern-.05em{\sc i\kern-.025em b}\kern-.08em
    T\kern-.1667em\lower.7ex\hbox{E}\kern-.125emX}}
\newlist{RQ}{enumerate}{1}
\setlist[RQ]{label=RQ\arabic*:}

\definecolor{Gray}{gray}{0.8}
\definecolor{LightGray}{gray}{0.9}
\newcolumntype{a}{>{\columncolor{LightGray}}c}
\newcolumntype{b}{>{\columncolor{white}}c}

\author{Ya Xiao$^*$, Salman Ahmed$^*$, Wenjia Song$^*$, Xinyang Ge$^\dagger$, Bimal Viswanath$^*$, Danfeng (Daphne) Yao$^*$}

\affiliation{%
  \institution{$^*$Computer Science, Virginia Tech, $^\dagger$ Microsoft Research}
}
\email{{yax99, ahmedms, wenjia7, vbimal, danfeng}@vt.edu, xing@microsoft.com}

\begin{document}

\title[Embedding Code Contexts for Cryptographic API Suggestion]{Embedding Code Contexts for Cryptographic API Suggestion: New Methodologies and Comparisons}

\begin{abstract}
 Despite recent research efforts, the vision of automatic code generation through API recommendation has not been realized. Accuracy and expressiveness challenges of API recommendation needs to be systematically addressed. We present a new neural network-based approach, Multi-HyLSTM for API recommendation --- targeting cryptography-related code. Multi-HyLSTM leverages program analysis to guide the API embedding and recommendation. By analyzing the data dependence paths of API methods, we train embeddings and specialize a multi-path neural network architecture for API recommendation tasks that accurately predict the next API method call. We address two previously unreported programming language-specific challenges, differentiating functionally similar APIs and capturing low-frequency long-range influences. Our results confirm the effectiveness of our design choices, including program-analysis-guided embedding, multi-path code suggestion architecture, and low-frequency long-range-enhanced sequence learning, with high accuracy on top-1 recommendations. We achieve a top-1 accuracy of 91.41\% compared with 77.44\% from the state-of-the-art tool SLANG. In an analysis of 245 test cases, compared with the commercial tool Codota, we achieve a top-1 recommendation accuracy of 88.98\%, which is significantly better than Codota's accuracy of 64.90\%. We publish our data and code as a large Java cryptographic code dataset.
\end{abstract}

\maketitle

\section{Introduction}


Modern software relies heavily on standard or third-party libraries. However, learning the correct usage of countless APIs is challenging for developers due to API complexity and insufficient documentation~\cite{kruger2017cognicrypt}, and often leads to misuse of APIs~\cite{meng2018secure, votipka2020understanding, acar2016you}. For example, multiple studies have shown that misusing Java cryptographic APIs is common in practice~\cite{egele2013empirical,rahaman2019cryptoguard,chen2019reliable}. Thus, a tool suggesting correct API methods would help developers from misuse.

Despite the significant progress in suggesting the next API(s), there is still room for improving prediction accuracy, especially for critical and complex APIs. Most state-of-the-art API recommendation solutions~\cite{raychev2014code, nguyen2016learning, nguyen2016api, svyatkovskiy2019pythia,nguyen2020code} utilize machine learning models with program analysis and suggest top-{\em k} APIs. 
We aim to design reliable API method recommendation approaches with high \textbf{top-1 accuracy}, as well as comprehensively compare the accuracy capabilities of multiple program analysis and programming language modeling methodologies. 

Making accurate recommendations on API method calls is challenging due to several reasons. \textit{First}, the programming context necessary to make a precise recommendation is often missing. In real-world codebases, many essential code dependencies often locate in other methods or files far away from the places of recommendation queries. Source code sequences, abstract syntax trees (ASTs), and \textit{intraprocedural} static analysis traces~\cite{raychev2014code,hindle2012naturalness,ray2016naturalness,chen2019sequencer} are likely to miss them due to the insufficient coverage beyond the local method. 

\begin{figure*}
    \centering
    \includegraphics[width=\textwidth]{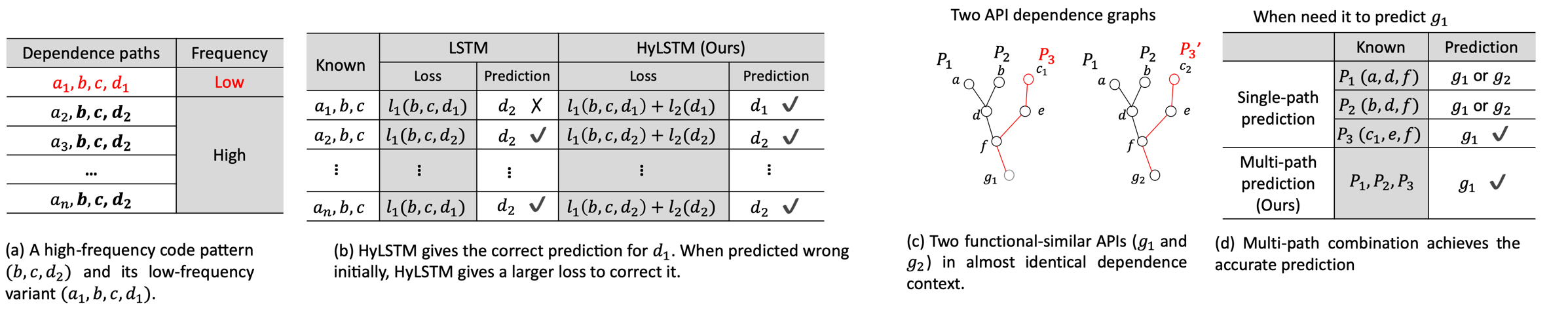}
    \caption{\small{Examples illustrating the two programming specific challenges and how we fix them.}}
    \vspace{-1em}
    \label{fig:example_frequency}
\end{figure*}

\textit{Second}, neural network models need to be specialized to address programming language-specific challenges. For example, we find that existing sequence models (e.g., LSTM) only capture the most frequent API sequence patterns. There are many low-frequency patterns that are neglected. This issue is more severe when a less frequent pattern shares a common suffix with a high-frequency one (Figure~\ref{fig:example_frequency} (a)). This shared segment misleads LSTM to make a wrong prediction.  
Recent Transformer models that use an attention mechanism~\cite{vaswani2017attention,devlin2018bert} to identify important parts of the input sequence, do not solve the frequency imbalance problem as well.


\sloppy{
\textit{Third}, another programming language-specific challenge is differentiating between functionally-similar, but non-interchangeable API methods.
For example, \lstinline|new String(byte[])| and \lstinline|Base64.Encoder.encodeToString(byte[])| are two candidate APIs after \lstinline|Cipher| decryption to restore the decrypted \lstinline|byte[]| to the original \lstinline|String|. However, the APIs are not interchangeable\footnote{In contrast, synonyms are usually interchangeable in natural language settings.}. 
We show that capturing multiple dependence paths can help for precise suggestions in such cases.
\par}

To address the above challenges, we present a neural network-based API recommendation technique called Multi-HyLSTM. Our approach achieves high top-1 accuracy, making it better suitable for real-world usage, compared to existing approaches. Our key insight is to build API embeddings guided by program analysis techniques, and to use neural network architectures that are better suited for programming languages. 

Multi-HyLSTM relies on program analysis to extract more complete dependence relations and represent them as dependence-aware embeddings. In this process, we construct API dependence graphs on program slices to specify data dependence paths between 
API methods. The dependence paths are used to train our dependence-aware embedding (i.e., vectorized representation of code elements) representing the API methods and associated constants. 
Multi-HyLSTM differs from existing sequential models as it incorporates multiple dependence paths reaching the query place to generate accurate recommendations.
Compared with the state-of-the-art sequential models (e.g., LSTM, BERT), our approach is better at capturing rare API patterns, as well as long-range complex dependencies. 

The API recommendation work closest to us is SLANG~\cite{raychev2014code}, which also relies on the combination of program analysis and deep learning approach. Compared to SLANG, we extract more complete dependence information. More importantly, we develop new machine learning techniques to recognize rare and complex API patterns. Our evaluation shows that Multi-HyLSTM outperforms SLANG in capturing more comprehensive dependencies and making more accurate top-1 recommendations.


We extensively evaluate our approach with Java cryptographic code extracted from Android apps. We choose Java cryptographic code because the associated APIs are notoriously complex and error-prone~\cite{chen2019reliable,afrose2019cryptoapi}. The evaluation includes two parts. First, we compare the top-1 accuracy of our Multi-HyLSTM with the state-of-the-art SLANG~\cite{raychev2014code} and a commercial IDE plugin Codota~\cite{codota}. 
Second, we conduct extensive comparative analysis to measure the effectiveness of each design choice via a new metric, \textit{in-set accuracy} (See Section~\ref{sec:eval_design}). This new metric measures accuracy of top-1 recommendations that fall in a plausible next API method set. We collect the next API method set for every input sequence from our dataset by considering the multiple available choices.  

Our key findings are summarized as follows.
\begin{itemize}
\item We evaluate the top-1 recommendation accuracy of Multi-HyLSTM on over 36K Java cryptographic API method invocations. Multi-HyLSTM achieves a top-1 accuracy of 91.41\%, a 14\% improvement over SLANG's~\cite{raychev2014code} accuracy of 77.44\%. We also conduct a manual analysis to compare Multi-HyLSTM with a state-of-the-art commercial IDE plugin Codota~\cite{codota} using 245 randomly selected test cases. Our method generates substantially more accurate API recommendations with a top-1 accuracy of 88.98\%, a 24\% improvement over Codota's top-1 accuracy of 64.90\%.


\item  Multi-HyLSTM achieves the best \textit{in-set accuracy} at 98.99\%. 
 It shows substantial accuracy improvement over BERT (from 55.73\% to 83.02\%) on previously unseen cases.


\item  We obtain the program-analysis guided embedding \textit{slice2vec} and \textit{dep2vec} for representing API methods and constants. 
Compared with the one-hot encoding on byte code that does not have program-analysis guidance, \textit{slice2vec} and \textit{dep2vec} both show substantial improvements on the \textit{in-set accuracy} by 28\%, 36\%, respectively.



\item 
 HyLSTM, the important building block of Multi-HyLSTM, achieves an excellent \textit{in-set accuracy} (93.00\%) on single dependence paths. HyLSTM shows 13.73\% higher \textit{in-set accuracy} on previously unknown cases over a regular LSTM model.

\end{itemize}

We also publish a large-scale Java cryptographic code dataset \footnote{ \url{https://github.com/Anya92929/DL-crypto-api-auto-recommendation}} for evaluating embedding and suggestion model designs. It includes the code corpora of bytecode, slices, and data dependence paths.



\begin{figure*}
    \centering
    \includegraphics[width=.8\linewidth]{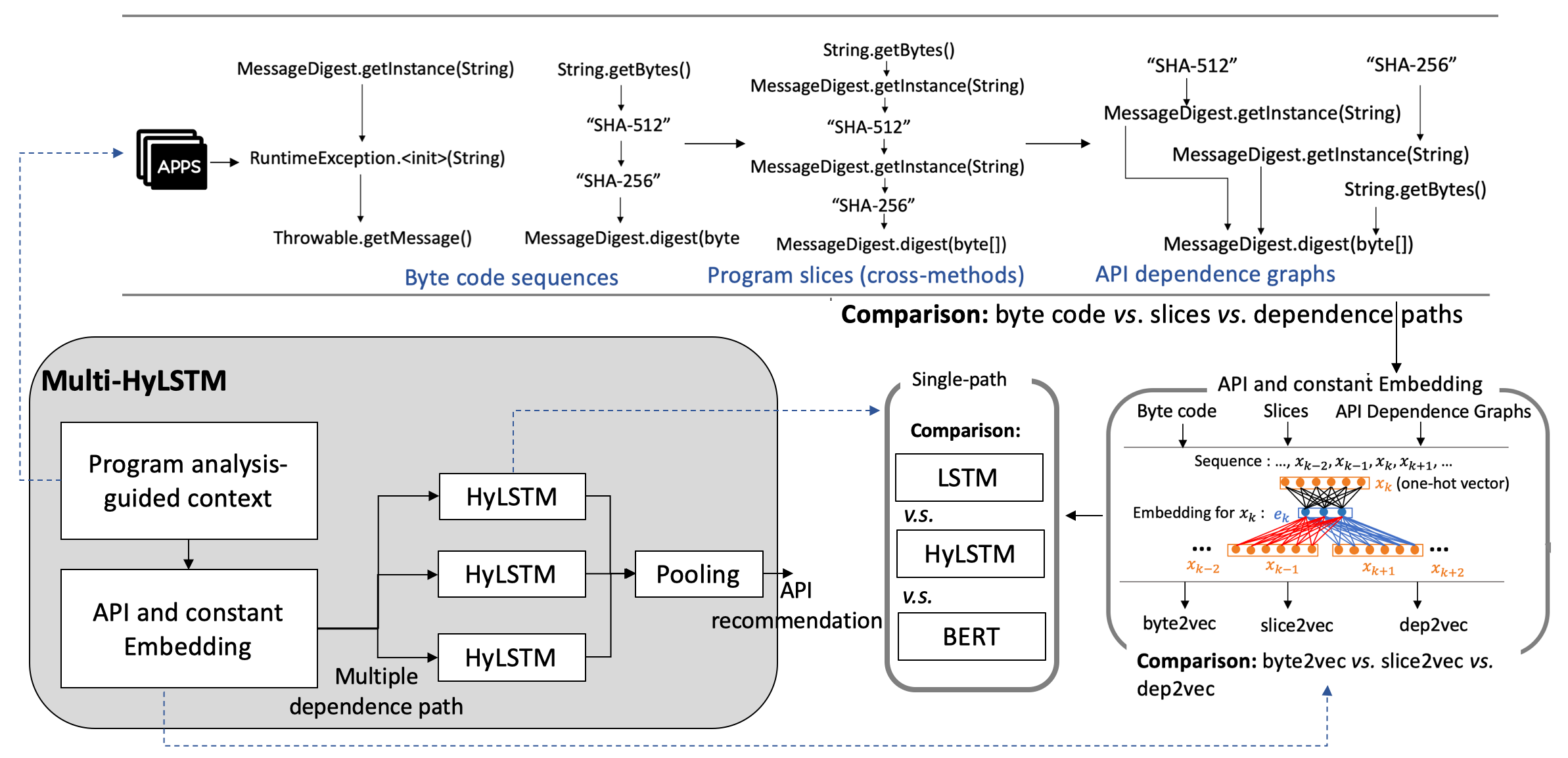}
    \vspace{-1.5em}
    \caption{\small{Main components of our API recommendation solution Multi-HyLSTM and the overview of our comprehensive comparative setting}}
    \label{fig:overview}
    \vspace{-1.5em}
\end{figure*}

\section{Technical Challenges and Research Questions}
\label{sec:challenges}


%

We discuss two programming language-specific challenges that hinder accurate API method recommendations. 
We use the term \textit{API pattern} to refer to the subsequence of API methods that frequently appear in the data-dependence paths. 
We call two API patterns sharing a common subsequence as the \textit{variant of each other}.

\smallskip
\noindent
{\em Design challenge 1: how to recognize the low-frequency variant of a high-frequency API pattern?}

Some API patterns are used more frequently than their variants, making it difficult to recognize their low-frequency variants. The end result is that the model only predicts the frequent API pattern.
An example is shown in Figure~\ref{fig:example_frequency}(a). The subsequence ($b, c, d_2$) is a high-frequency pattern in our extracted data-dependence paths. However, when the prefix is a special one (e.g., $a_1$), the last API method should be $d_1$, instead of $d_2$. This phenomenon is common in real codebases. 
We find that native LSTM models are unable to identify the low-frequency variant (i.e., $a_1, b, c, d_1$) because of the frequency imbalance. 


To address this issue, we present a new sequential model HyLSTM by modifying the LSTM loss function (Section~\ref{sec:hylstm}). Our idea is to give a strong signal to the model about the different patterns. We amplify the importance of the last token (e.g., node $d_1$ and $d_2$) in the entire sequence to calculate the sequence loss. Our experiments verify that this new amplified loss function in HyLSTM makes the model better at identifying low-frequency variants. We use HyLSTM as an important building blocks of our Multi-HyLSTM.

\smallskip
\noindent
{\em Design challenge 2: how to differentiate different APIs that share similar functionality?} 

In Java APIs, even though slightly different APIs may share the same functionality, they cannot be used interchangeably. The prediction model needs to know when to use which API. Existing work does not address this challenge. In Figure~\ref{fig:example_frequency} (c), the API methods $g_1$ and $g_2$ work in a similar programming context. According to the dependence paths $P_1$ and $P_2$, there is no indicator showing which one should be used. To distinguish them, a critical path $P_3$ should be captured. We find that many functionally-similar API methods follow this property.   
For example, \lstinline|new String(byte[])| and \lstinline|Base64.Encoder.encodeToString(byte[])| both work for encoding \lstinline|byte| arrays into a \lstinline|String| after the identical \lstinline|Cipher| decryption operations. The shared dependence paths make them indistinguishable. 
Our approach is to incorporate all the dependence paths from their arguments to increase the coverage for the critical dependencies (e.g., previous decoding behaviors for this example). Our recommendation is therefore made by leveraging multiple dependence paths.

Our work also systematically compares various design choices for neural-network-based API recommendations. We make substantial efforts to confirm the effectiveness of a model design through comparative experiments. Our research questions are: 

\begin{itemize}
    \item RQ1: Does our approach outperform the state-of-the-arts in terms of the top-1 recommendation?
    \item RQ2: How  much  do  program-analysis  insights  improve  API recommendation?
    \item RQ3: How  to  recognize  low  frequency  long  API  sequences?
    \item RQ4: How  to  accurately  differentiate  different APIs that share the same functionality?
\end{itemize}


\section{Design of Multi-HyLSTM }

We describe the key design choices of our Multi-HyLSTM. We start by introducing our dependence-aware embedding guided by static program analysis. The static analysis draws the data dependence paths that are required to make accurate API recommendations. Based on the extracted dependence paths, embedding maps thousands of cryptography-related API methods and constants to the dependence-aware vectors of a low dimension (e.g., 300). We then describe the new neural network model designed for high-precision API recommendation, including the multi-path architecture and a low-frequency long-range-enhanced model called HyLSTM.
%
%
The main components of our approach and the overview of our comparative analysis are illustrated in Figure~\ref{fig:overview}.




\subsection{Program-analysis-guided Embedding}\label{sec:embedding}

We specialize lightweight static data-flow analysis techniques to preprocess code sequences before embedding them. The embedding of API methods and constants act as the input vectors of our neural network model. By applying the static analysis, our approach aims to effectively leverage programming language insights to capture dependence relations in embedding vectors.


\noindent
{\sc Inter-procedural backward slicing.} 
   We start the inter-procedural backward slicing from the invocation statement of a cryptography API. The slicing criteria are its arguments. All the code statements influencing the API invocation are organized together to form a program slice. Self-defined method calls are replaced by their code body statements. By eliminating the non-influencing code statements and breaking the method boundary, the program slices guarantee the required information program-wide as well as keeping the code size minimum. It also partially disentangles the API and constant sequences as shown in Fig.~\ref{fig:overview}.

\noindent
{\sc API dependence graph construction.}
On the program slice, we build the API dependence graph by adding data dependence edges on its control flow graph. Then, we simplify this graph as a high-level API dependence graph (see Fig.~\ref{fig:overview}), in which each node is an API or a constant, by merging the non-API nodes. It significantly reduces the size of a graph and connects two data-dependent APIs. 





We apply skip-gram embedding model~\cite{mikolov2013distributed} on the byte code, slices, and dependence paths extracted from the API dependence graphs, respectively, to produce three types of Java cryptography embeddings, \textit{byte2vec}, \textit{slice2vec}, and \textit{dep2vec}.  
Figure~\ref{fig:overview} gives simple examples.

\begin{itemize}
    \item \textit{byte2vec} is the baseline embedding version that applies \textit{word2vec} \cite{mikolov2013efficient,mikolov2013distributed} directly on the byte code corpus.
    \item  \textit{slice2vec} is the embedding with the inter-procedural backward slicing as the pre-processing method.
    \item  \textit{dep2vec} applies two pre-processing techniques, inter-procedural backward slicing, and API dependence graph construction, to guide the embedding training. 
\end{itemize}


\begin{figure}[b]
\vspace{-1em}
    \centering
    \includegraphics[width=\linewidth]{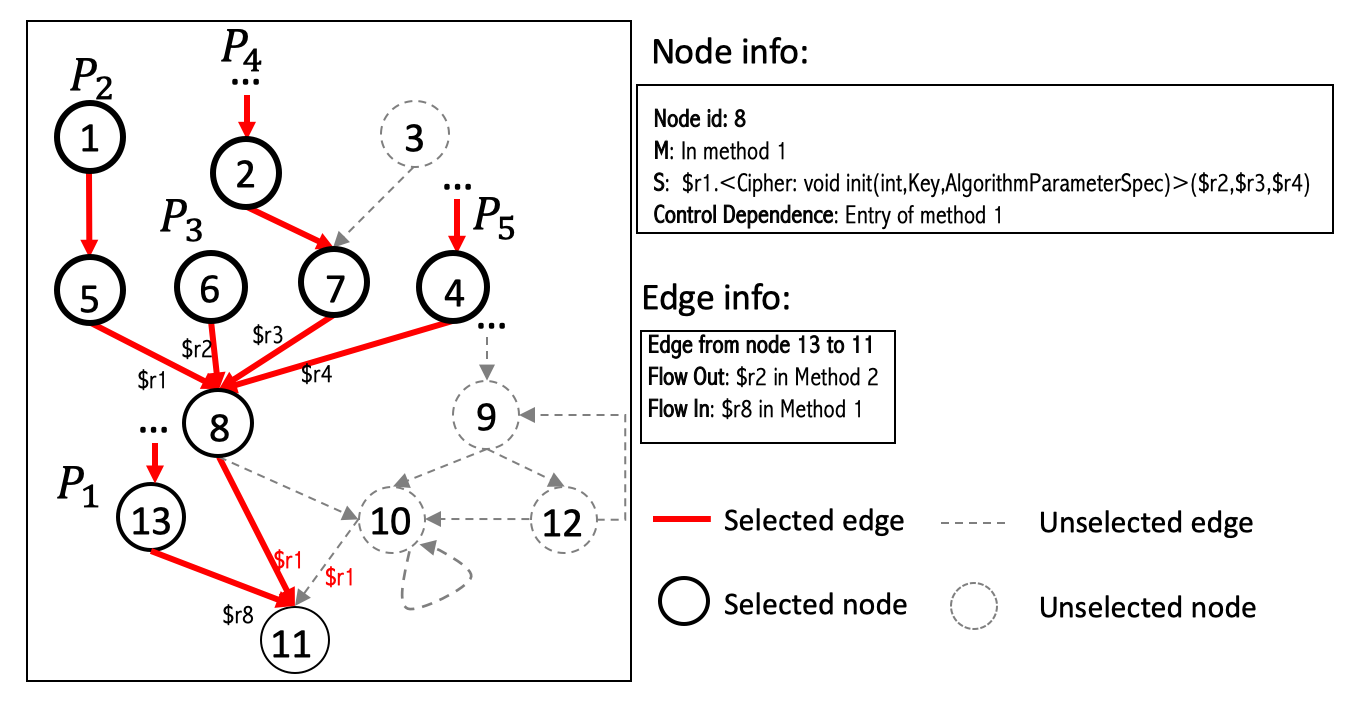}
    \caption{\small{Multiple Paths Selection for API Recommendation. We use the information associated with the nodes and edges to select paths. The goal is to maximize the coverage for different nearby branches with minimal paths.}}
    \label{fig:path_selection}
    \vspace{-1em}
\end{figure}
\subsection{Multi-path API Recommendation}\label{multi_hyLSTM}



We design a new multi-path architecture and apply it on top of the sequential language models to address the challenge caused by functionally-similar-yet-non-interchangeable APIs. Instead of making API recommendations based on a single sequence, our new model incorporates multiple related data dependence paths. This design allows the model to have a more comprehensive view of the data flows, thus producing more accurate predictions.  The training of multi-path API recommendation model includes three operations: {\sc multi-path selection}, {\sc path-embedding training}, and {\sc multi-path aggregation}.
The multi-path architecture leverages the efficiency and performance success achieved by sequence models that are more efficient than the graph-based neural networks that suffer the scalability issue.

\smallskip
\noindent
{\sc Multi-path Selection.} 
The number of paths in an API dependence graph sometimes explodes. To minimize the computational cost, we apply our multi-path selection algorithm to limit the number of paths (to five in our experiments) that are input to neural networks. Intuitively, our goal is to \textbf{maximize} the coverage of different nearby data and control flow branches with a \textbf{minimal} number of paths. Specifically, we traverse the paths backward from the target node to the beginning node. At each branch, we apply the breadth-first search to select the edges that deliver different flow-in variables or belong to different control flow branches. 

As illustrated in
Figure~\ref{fig:path_selection},
there are flow-out and flow-in variables associated with each edge and control dependence information associated with each node. 
For node 11, there are two edges from node 8 and node 10, respectively, delivering identical flow-in variables $\$r1$. One can select either one of them. In this example, node 8 and node 13 that deliver different variables are selected to continue the path traversal.  
If there are two predecessors that deliver identical flow-in variables but belong to different control flow branches, we also add both of them in this step. We continue this breadth-first backward traversal process until the path budget has been used up. After that, each selected branch will be completed as a complete path by the depth-first search to an arbitrary beginning node.  This greedy breadth-first approach guarantees the local optimal choice at every branch. Since the nearby dependencies are more important than the faraway ones, our backward traversal prioritizes the coverage of the nearby dependencies. 


\smallskip
\noindent
{\sc Path Embedding Training.} To better understanding each path, we generate the sequence embedding, the vectors for an entire sequence, by pretraining the language model with all the separate dependence paths extracted from the API dependence graphs. The language model takes these sequences as input and is trained at every step to predict the next token. Every token of this sequence is represented as the \textit{dep2vec} embedding we trained for each cryptographic API and constants in Section~\ref{sec:embedding}. We use the hidden state vector after the last timestep as the embedding for the input sequence. This vector describes the probability distribution of the next API usage given this path $P(t_n|(t_0,t_1,\dots,t_{n-1}))$. This sequence model is fine-tuned later under the multi-path architecture. 

\begin{figure}
    \centering
    \includegraphics[width=0.8\linewidth]{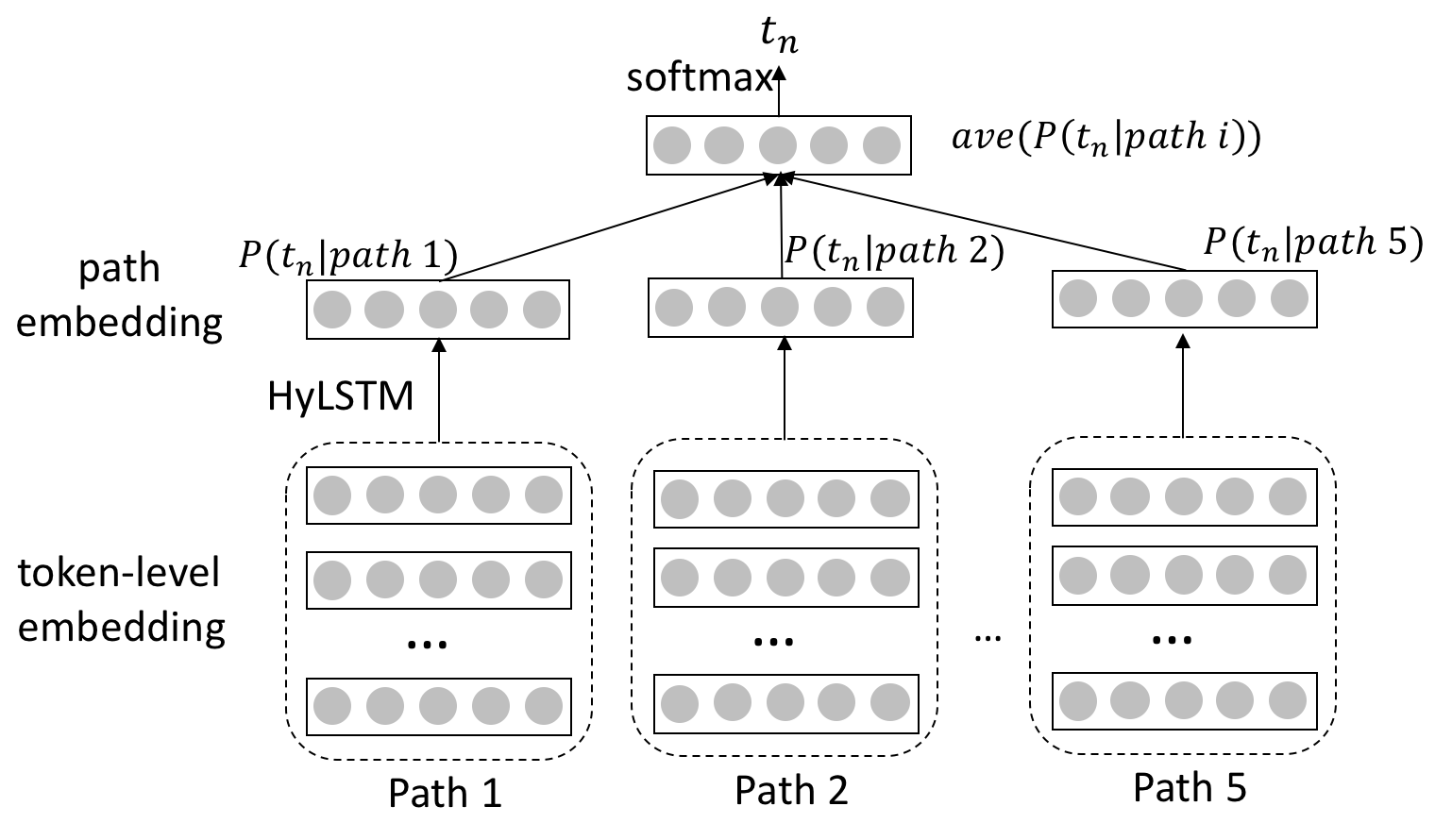}
    \caption{\small{Multi-path Code Suggestion based on aggregating the Path Embedding}}
    \label{fig:multi_path_suggestion}
    \vspace{-1.5em}
\end{figure}

\smallskip
\noindent
{\sc Multi-path Aggregation.} Based on the selected multiple dependence paths, we predict the next API by aggregating their sequence embedding vectors. As shown in Figure~\ref{fig:multi_path_suggestion}, the pretrained sequence model is set as the initial state.  Each of them produces path embedding that suggests the next token $t_n$ according to $P(t_n|(t_0,t_1,\dots,t_{n-1}))$. We add an average pooling layer to aggregate these path embeddings into one vector that represents $ave(P(t_n|path\ i))$. The prediction is generated from that aggregated vector with the softmax classifier. Under the multi-path architecture, the sequence model is jointly updated with the following layers towards the task-specific distribution.  The multi-path architecture successfully highlights the minor difference in the process of code suggestion.   Candidate tokens having similar probabilities in some paths can be accurately differentiated based on their probability difference from an extra path. We applied this multi-path architecture on the sequence model BERT and HyLSTM, an advanced LSTM model proposed by us, in Section~\ref{sec:multi-path} to demonstrate the effectiveness of our approach.

\subsection{Low-frequency Long-range-enhanced Learning}\label{sec:hylstm}

%
\begin{figure}
    \centering
    \includegraphics[width=.66\linewidth]{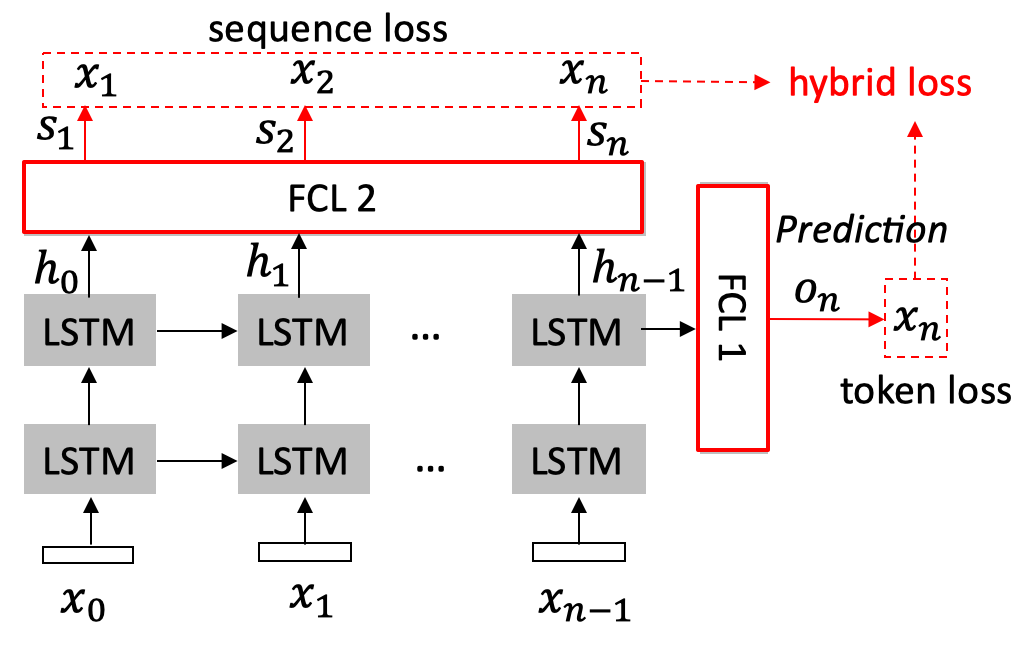}
    \caption{\small{Target amplification in HyLSTM with our new hybrid loss function.}}
    \label{fig:hylstm}
    \vspace{-1.5em}
\end{figure}

Under the multi-path architecture, each dependence path is handled by a sequence model presented by us, referred to as HyLSTM. HyLSTM is designed for addressing the challenge posed by the low-frequency API method sequences with a high-frequency suffix, as we demonstrate in Figure~\ref{fig:example_frequency}. Intuitively, we solve this issue by forcing the model to assign larger weights to beginning tokens (e.g., $a_1$ in Figure~\ref{fig:example_frequency}) when needed, making beginning tokens more useful for predicting the API (e.g., $d_1$).

HyLSTM differs from the regular LSTM sequence model in its architecture and loss function. 
We show our HyLSTM in Figure~\ref{fig:hylstm}. It includes two parallel output layers $FCL_1$ and $FCL_2$ after the LSTM cells while regular LSTM based sequence learning only has the $FCL_2$.  The output of $FCL_1$ is the target (i.e., the recommended API method) of the given dependence path while the $FCL_2$ outputs every intermediate token of the given path.  


As regular LSTM, the hidden states $h_i$ at every timestep are taken by the fully connected layer  $\mbox{\em FCL}_2$ to output the next token in the forward propagation. 
\begin{align}
\begin{bmatrix}
s_1\\ 
s_2\\ 
\dots\\ 
s_{n}
\end{bmatrix}= \mbox{\em softmax}(\begin{bmatrix}
h_0\\ 
h_1\\ 
\dots\\ 
h_{n-1}
\end{bmatrix} W_2 +\begin{bmatrix}
B_2\\ 
B_2\\ 
\dots\\ 
B_2
\end{bmatrix})
\end{align}
where $W_2$ and $B_2$ are the weights and bias for $\mbox{\em FCL}_2$.

In HyLSTM, we add an extra parallel fully connected layers $\mbox{\em FCL}_1$ to accept the hidden layer at the last timestep $h_{n-1}$. $\mbox{\em FCL}_1$ outputs the prediction $o_{n}$ by:
\begin{align}
 o_{n} = \mbox{\em softmax}( h_{n-1} W_1 + B_1)
 \end{align}
 where $W_1$ is a weight matrix and $B_1$ is the bias vector for $\mbox{\em FCL}_1$.
 
 We define a hybrid loss combining the losses from the two fully connected layers. The goal of this design is to guarantee that the hidden states $h_{n-1}$ are distinguishable to reproduce the different targets via $FCL_1$, as well as maintain the similarity at the intermediate steps via $FCL_2$.
 

The new hybrid loss $l_h$ is defined as:
\begin{align}
l_h = \alpha  L(o_n,x_n)+ (1-\alpha) \sum_{i=1}^{n}\frac{L(s_i,x_i)}{n}
\end{align}
where $L(\cdot)$ is the cross entropy loss between the output and label. We set  $\alpha$ to be 0.5 in our experiments.


When low-frequency sequences were initially predicted wrong due to its misleading high-frequency suffix, HyLSTM produces a larger loss than regular LSTM to correct it and treat the beginning tokens more seriously.
Besides evaluating HyLSTM against regular LSTM models, we also compare it with BERT in Section~\ref{sec:experimental_results}.


\smallskip
\noindent
{\em Integration of multi-path architecture and HyLSTM.}
We apply our multi-path architecture on top of HyLSTM to make API recommendations with high accuracy, which produces our Multi-HyLSTM model. 




\section{Datasets and Experimental Setups}\label{sec:dataset_setup}



We collect two datasets of 36,641 and 12,335 apps from the Google Play store, respectively. 
To build a diverse dataset, we select the most popular apps from different categories in the app store and filter them to guarantee that these  apps use Java cryptographic APIs. 
The former dataset is used to train and evaluate the API method and constant embedding. The latter is used for downstream tasks training and evaluation.  




To build embeddings, the following three types of code corpora are extracted during pre-processing: {\em i) byte code}, {\em ii) slices}, and {\em iii) dependence paths}. Their size is shown in Table~\ref{tab:stats_corpora}. The tokens include the APIs or constants we aim to embed. APIs include the standard APIs from Java and Android platforms, as well as some third party APIs that cannot be inlined because they are recursive or phantom (whose bodies are inaccessible during the analysis). We filter APIs to embed those that appear more than 5 times. Besides, we manually collect 104 reserved string constants used as the arguments of cryptographic APIs. The 104 constants along with the constants that appear more than 100 times in the slice corpus are used for embedding. Finally, we have 4,543 tokens (3,739 APIs and 804 constants) in the embedding vocabulary.

\begin{table}[t]
\centering
\caption{\small{Statistics of embedding corpora size}}
\label{tab:stats_corpora}

\begin{tabular}{baba}
\Xhline{2\arrayrulewidth}
\rowcolor{Gray}
\textbf{Corpora} & \textbf{Byte code} & \textbf{Slices} &  \begin{tabular}[c]{@{}c@{}}\textbf{Dependence}\\ \textbf{paths}\end{tabular}  \\ \hline
\textbf{\# of tokens} & 28,887,852 & 12,341,912 & 38,817,046 \\ \Xhline{2\arrayrulewidth}
\end{tabular}
\vspace{-0.5em}
\end{table}

\begin{table}[ht]
\centering
\caption{\small{Composition of the embedded APIs}}
\label{tab:composition_apis}

\begin{tabular}{lla}

\Xhline{2\arrayrulewidth}
\rowcolor{Gray}
\multicolumn{2}{c}{\textbf{Source}} & \multicolumn{1}{l}{\textbf{\# of embedded APIs}} \\ \hline
\hline
\multicolumn{1}{l|}{\multirow{5}{*}{Java platform}} & \multicolumn{1}{l|}{java.security} & 510 \\
\multicolumn{1}{l|}{} & \multicolumn{1}{l|}{javax.crypto} & 166 \\
\multicolumn{1}{l|}{} & \multicolumn{1}{l|}{java.io} & 138 \\
\multicolumn{1}{l|}{} & \multicolumn{1}{l|}{java.lang} & 259 \\
\multicolumn{1}{l|}{} & \multicolumn{1}{l|}{others} & 374 \\ \hline
\multicolumn{2}{l|}{Android platform} & 486 \\ \hline
\multicolumn{2}{l|}{Third parties} & 1827 \\ \Xhline{2\arrayrulewidth}
\end{tabular}
\vspace{-1em}
\end{table}



Table~\ref{tab:naturalness_of_data} shows the number of sequences extracted from different code corpora.
Compared with byte code, slices and dependence paths have more sequences, but less \textbf{unique} sequences.
This suggests that learning under the guidance of program-analysis is more desirable than learning on byte code.
To guarantee fairness, we subsample the extremely frequent dependence paths as \cite{mikolov2013distributed} without losing the unique paths. The total number of paths after subsampling is 566,279, which has a close total/unique ratio with slices. 

\begin{table}[b]
\centering
\vspace{-1em}
\caption{\small{Number of sequences parsed from different representations}}
\vspace{-1em}
\begin{tabular}{|b|a|b|a|}
\hline \hline
\rowcolor{Gray}
\textbf{}       & \textbf{\begin{tabular}[c]{@{}c@{}}Byte\\ code\end{tabular}} & \textbf{Slices} & \textbf{\begin{tabular}[c]{@{}c@{}}Dependence\\  paths\end{tabular}} \\ \hline  \hline
\textbf{Total}  & 707,775                                                         & 926,781          & 4,655,763                                                              \\ \hline
\textbf{Unique} & 262,448                                                         & 54,173           & 23,128                                                                \\ \hline  \hline
\end{tabular}
\vspace{-1.5em}
\label{tab:naturalness_of_data}
\end{table}

\smallskip
\noindent
{\em Experimental setup of program analysis pre-processing}.
To get slices, we develop a crypto slicer that is inter-procedural, context- and flow-sensitive. We use the open-source Java program analysis framework Soot~\cite{vallee2010soot} for pre-processing. Soot takes the Android byte code as input and transforms it into an intermediate representation (IR) Jimple. The program analysis is performed on Jimple IR. Our slicer using Soot's BackwardFlowAnalysis framework works with Soot 2.5.0, Java 8, and Android SDK 26.1.1.

\smallskip
\noindent
{\em Experimental setup for embedding}. 
We use the same hyperparameters decided by our preliminary experiments for all the embedding training. The embedding vector length is 300. The sliding window size for neighbors is 5. We also applied the subsampling and negative sampling to randomly select 100 false labels to update in each batch. We train embeddings with the mini-batch size of 1024. The embedding terminates after 10 epochs. 
Our embedding model is implemented using Tensorflow 1.15. Training runs on the Microsoft AzureML GPU clusters, which supports  distributed training with multiple workers. We use a cluster with 8 worker nodes. The VM size for each node is the (default) standard NC6.

\smallskip
\noindent
{\em Experimental setup for API recommendation tasks}.
%
We filter the vulnerable code using CryptoGuard~\cite{rahaman2019cryptoguard}, which is a cryptography API misuse detection tool based on program analysis. Only secure code is kept to avoid data poisoning~\cite{schuster2020you}. 
We limit the maximum number of input steps to 10 (truncate sequences longer than 10). 
We use batch size of 1,024, and learning rate of 0.001. We select the highest accuracy achieved within 10 epochs because we observe that the accuracy increase is negligible after that. 
We use the stacked LSTM architecture with vanilla LSTM cells for the LSTM-based models. For the BERT model, we use 12 attention headers and the hidden size is 256.

  \begin{table*}[!b]
\caption{\small{The top-1 recommendation accuracy of Codota plugin and our Multi-HyLSTM on 245 randomly selected Java Cryptographic API invocation test cases. Codota gives recommendation based on the previous code and the return value while we only assumes that the previous code is known. We show our default accuracy without the return value and the improved version when the return value type is considered.} }\label{tab:codota}
\begin{small}
\begin{tabular}{|b|a|b|a|b|a|}
\hline
\hline
\rowcolor{Gray}
                                                        &  &  &  & \multicolumn{2}{c|}{\textbf{Multi-HyLSTM (Our approach) Accuracy}}  \\  \cline{5-6}  
\rowcolor{Gray} \multirow{-2}{*}{\textbf{App Category}} &  \multirow{-2}{*}{\textbf{\# of Test Cases}}                                    &            \multirow{-2}{*}{\textbf{\# of Apps}}                          &                     \multirow{-2}{*}{\textbf{Codota Accuracy}}                    & w/o return value         & with return value        \\ \hline \hline
Business                               & 66                                         & 3                                    & 66.67\%                                   & 89.39\%                  &    98.49\%                      \\ \hline
Finance                                & 99                                         & 3                                    & 65.66\%                                   & 90.91\%                  & 97.98\%                         \\ \hline
Communication                          & 80                                         & 3                                    & 62.5\%                                    & 86.25\%                  &   97.59\%                       \\ \hline
\textbf{Total}                                  & \textbf{245}                               & \textbf{9}                           & \textbf{64.9\%}                           & \textbf{88.98\%}         & \textbf{97.96\%}                \\ \hline \hline
\end{tabular}
\end{small}
\vspace{-1em}
\end{table*}

\section{Experimental Evaluation}\label{sec:experimental_results}

Our evaluation includes two parts. First, we evaluated the effectiveness of our Multi-HyLSTM with top-1 recommendation accuracy. We compare our accuracy results with the state-of-the-art tool SLANG~\cite{raychev2014code} and a commercial IDE plugin Codota~\cite{codota}. Second, we conducted extensive comparative experiments to analyze the accuracy capability of each design choice shown in Table~\ref{tab:comparative_setting}. The objective of this part is to study how effective these design choices are in terms of improving accuracy compared with some intermediate solutions and answer a couple of research questions. 

\noindent
\textbf{Evaluation Metrics.}
There are two metrics in our evaluation. We use the top-1 accuracy to evaluate our approach Multi-HyLSTM and two state-of-the-art tools SLANG and Codota on recommending the next API method. When evaluating the design choices (e.g., embedding vs. one-hot encoding, HyLSTM vs. LSTM) through the intermediate solutions, we introduce a new metric, referred to as \textit{in-set accuracy}.  
We use this new metric because we noticed that in the intermediate solutions, a code sequence (e.g., a dependence path) may have multiple correct choices for the next API method call. Therefore, we define \textit{in-set accuracy} as the accuracy of top-1 recommendations that fall in a reasonable next API method set. We collect the next API method set for every input sequence from our dataset by considering the multiple available choices. Compared with the top-$k$ accuracy, \textit{in-set accuracy} guarantees that the top-1 recommendation is a reasonable choice. 

\begin{figure}[t]
    \centering
    \includegraphics[width=.8\linewidth]{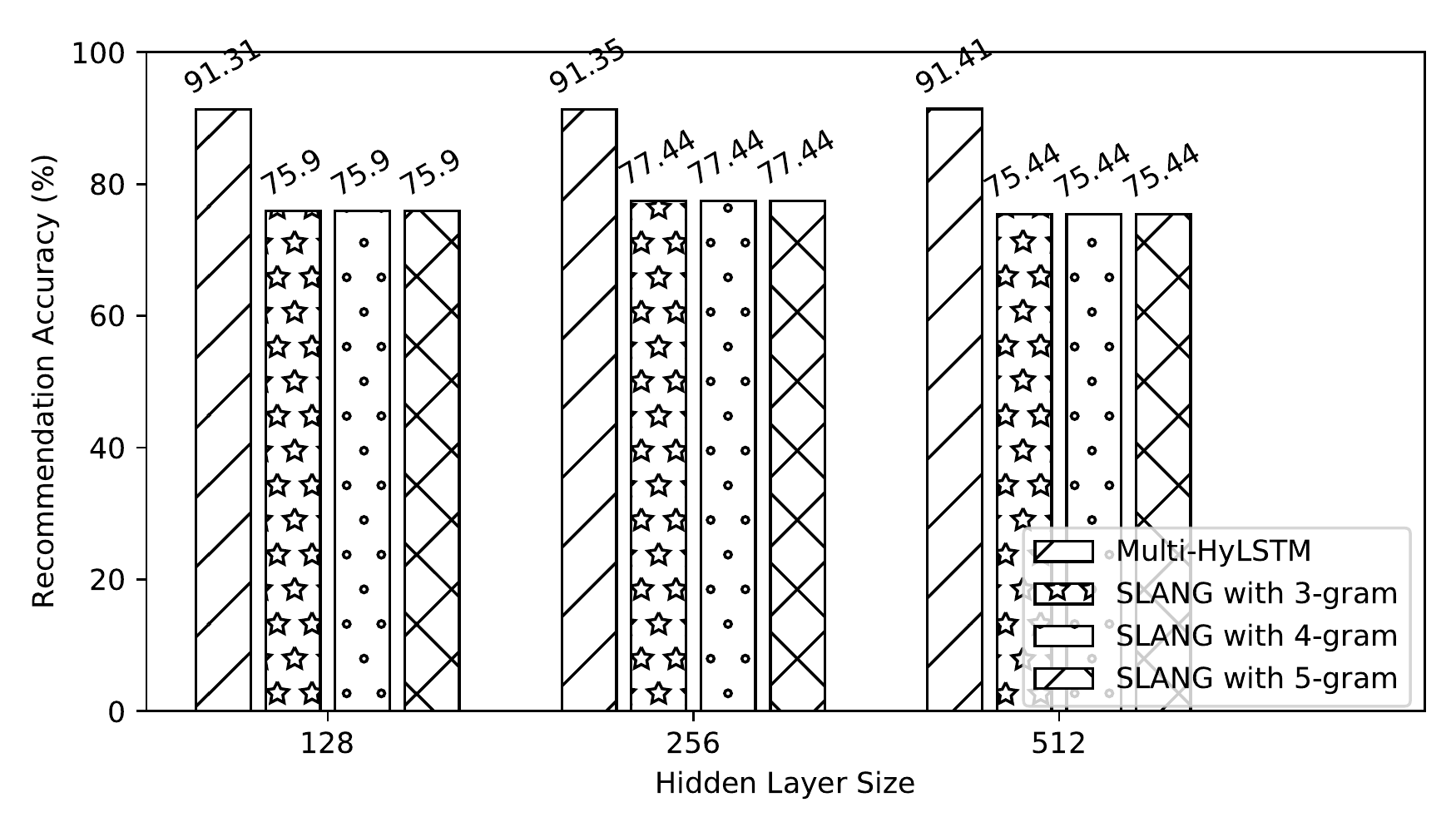}
    \vspace{-1em}
    \caption{\small{The top-1 accuracy of SLANG and our approach on recommending Java cryptographic API method calls from Android Applications }}~\label{fig:slang}
   \vspace{-1.5em} 
\end{figure}

\subsection{Top-1 Recommendation Accuracy}\label{sec:recom_eval}


\sloppy{
We perform experiments to answer our first research question (\textit{RQ1: Does our approach outperform the state-of-the-arts in terms of the top-1 recommendation?}). 
 We select two state-of-the-art API recommendation tools, SLANG~\cite{raychev2014code} and Codota~\cite{codota}, for comparison. SLANG completes the next API method based on program analysis and statistical language models. Codota is a commercial AI code completion plugin, which is adopted by the mainstream IDEs including IntelliJ, Eclipse, Android Studio, VS Code, etc. \par}

\noindent
\textbf{Comparison with SLANG}
 We locate 36,029 Java cryptographic API method callsites from the bytecode of the 16,048 Android Applications. By applying the required static analysis, our method and SLANG extract corresponding program context for these cryptographic API method callsites. We randomly select 1/5 of the callsites as the test cases and train SLANG and our model with the other 4/5 cases.    
SLANG combines the $n$-gram and RNN models to generate the probability of the next API method call. 
We try different $n$ for the $n$-gram model and different hidden layer sizes for the RNN model of SLANG and our Multi-HyLSTM model.     

Figure~\ref{fig:slang} shows the top-1 accuracy of SLANG and our approach under different configurations. We choose the hidden layer size of SLANG and our approach from 128, 256, and 512. With each hidden layer size setting, we also adjust SLANG with 3-gram, 4-gram, 5-gram model~\footnote{Raychev et al choose 3-gram and hidden layer size 40 for RNN in \cite{raychev2014code}}. Our approach shows significant advantages over SLANG under all settings. The highest top-1 accuracy of SLANG is 77.44\%, achieved with RNN-256. The $n$-gram model shows no impact on the top-1 accuracy. Our models achieves the best accuracy at 91.41\% under Multi-HyLSTM with hidden layer size 512.

\noindent
\textbf{Comparison with Codota}
Given an incomplete code statement with an object with a \verb|dot| in an IDE, Codota displays a ranked list of the recommended API methods associated with the object. We randomly selected 245 cryptographic API method invocations from 9 Android applications as the test cases. We decompiled the 9 apps into source code and load them into IntelliJ IDE with Codota. Then, we manually triggered Codota recommendation by removing the method name after the \verb|dot|. The accuracy results are shown in Table~\ref{tab:codota}.  Our approach has a significant improvement on the top-1 accuracy compared with Codota. The top-1 accuracy of the 245 cases is improved from 64.90\% to 88.98\%. 

Codota takes not only the previous code but also the return value into consideration to give compatible recommendations. This is different from our approach which only relies on the previous code. 
For fair comparison, we add an extra manual evaluation on all these 245 test cases. We manual checked the return value type and use it to filter our recommendation candidates output from the neural network model.  Our top-1 accuracy (last column in Table~\ref{tab:codota}) rises to 97.96\% if the return type is known.

\subsection{Evaluation on Design Choices}\label{sec:eval_design}

\begin{table*}[htpb]
\vspace{-1em}
\caption{\small{The comparison settings of our comprehensive comparative analysis. Group 1 controls on our program analysis and embedding techniques. Group 2 focuses on the effectiveness of our low-frequency long-range enhancing technique. Group 3 reveals the improvement from the multi-path architecture. }}\label{tab:comparative_setting}
\vspace{-1em}
\begin{scriptsize}
\begin{tabular}{b|ababab|ab|abab}
\hline \hline
\rowcolor{Gray}  &\multicolumn{6}{c|}{Group 1 (RQ2)} & \multicolumn{2}{c|}{Group 2 (RQ3)} & \multicolumn{4}{c}{Group 3 (RQ4)}
\\ \cline{2-13}
    \rowcolor{Gray}              Our design choices                                       & \begin{tabular}[c]{@{}c@{}}1-hot on\\ bytecode\end{tabular} & byte2vec & \begin{tabular}[c]{@{}c@{}}1-hot on\\ slices\end{tabular} & slice2vec & \begin{tabular}[c]{@{}c@{}}1-hot on\\ dependence \\ paths\end{tabular} & dep2vec & HyLSTM & \begin{tabular}[c]{@{}c@{}}LSTM \\ (token-level or \\ sequence-level loss)\end{tabular} & \begin{tabular}[c]{@{}c@{}}HyLSTM\\ (path embedding)\end{tabular} & BERT & Multi-HyLSTM & Multi-BERT \\ \hline \hline
\begin{tabular}[c]{@{}c@{}}Interprocedural \\ program slicing\end{tabular}   & \emptycirc                                                           & \emptycirc        & \fullcirc                                                 & \fullcirc     & \fullcirc                                                                 & \fullcirc    & \fullcirc   & \fullcirc                                                             & \fullcirc                                        & \fullcirc                        & \fullcirc       & \fullcirc     \\ \hline
\begin{tabular}[c]{@{}c@{}}API dependence \\ graph construction\end{tabular} & \emptycirc                                                   & \emptycirc        & \emptycirc                                                   & \emptycirc     & \fullcirc                                                               & \fullcirc  & \fullcirc   &\fullcirc                                                           & \fullcirc                                           & \fullcirc                       &\fullcirc       & \fullcirc      \\ \hline
\begin{tabular}[c]{@{}c@{}}Token-level\\ embedding\end{tabular}                                                                    & \emptycirc                                                         & \fullcirc    &\emptycirc                                                   & \fullcirc      & \emptycirc                                                               & \fullcirc   & \fullcirc     & \fullcirc                                                               & \fullcirc        & \fullcirc                                                           & \fullcirc          & \fullcirc        \\ \hline
\begin{tabular}[c]{@{}c@{}}Low frequency long\\ range enhancing\end{tabular} & \emptycirc                                                        & \emptycirc     & \emptycirc                                                 & \emptycirc       & \emptycirc                                                               & \emptycirc   & \fullcirc & \emptycirc    & \fullcirc                                                             & \emptycirc                                                              & \fullcirc          & \emptycirc     \\ \hline
\begin{tabular}[c]{@{}c@{}}Path-level\\ embedding\end{tabular}                                                     & \emptycirc                                                   & \emptycirc     & \emptycirc                                                   & \emptycirc     & \emptycirc                                                               & \emptycirc  & \emptycirc   & \emptycirc                                                            & \fullcirc                                        & \fullcirc                         & \fullcirc        & \fullcirc     
\\ \hline
\begin{tabular}[c]{@{}c@{}}Multi-path\\ architecture\end{tabular}                                                     & \emptycirc                                                   & \emptycirc     & \emptycirc                                                   & \emptycirc     & \emptycirc                                                               & \emptycirc  & \emptycirc   & \emptycirc                                                            & \emptycirc              & \emptycirc                                                    & \fullcirc        & \fullcirc       \\ \hline \hline
\end{tabular}
\end{scriptsize}

\end{table*}
We evaluate the effectiveness of our design choices  by comprehensive comparative experiments between the intermediate solutions. The intermediate solutions apply different combinations of our design choices. We demonstrate our comparative setting in Table~\ref{tab:comparative_setting}. 

\subsubsection{RQ2: How  much  do  program-analysis  insights  improve  API recommendation?}
We answer this research question with the comparative experiments in the group 1 of Table~\ref{tab:comparative_setting}. 

We compare these settings with two types of API recommendation tasks, i.e., {\em next API recommendation} and {\em next API sequence recommendation}. The former aims to predict one API method to be used in the next line while the latter produces a sequence of API methods to be invoked sequentially. 
We use the regular LSTM based sequence model for the next API method task. In the next API sequence recommendation task, we use the LSTM based seq2seq (encoder-decoder) model to accept the first half API element sequence and predict the last half API elements. 

\begin{table}[b]
\vspace{-1.5em}
\centering
\caption{\small{Accuracy of next API Recommendation.}}
\begin{scriptsize}
\vspace{-1em}
\begin{tabular}{|c|a|b|a|b|a|b|}
\hline \hline
\rowcolor{Gray}
\multirow{2}{*}{\textbf{\begin{tabular}[c]{@{}c@{}}LSTM\\ Units\end{tabular}}} & \multicolumn{2}{c|}{\textbf{Byte Code}} & \multicolumn{2}{c|}{\textbf{Slices}} & \multicolumn{2}{c|}{\textbf{Dependence Paths}} \\ \cline{2-7} 
                                                                               & 1-hot               & byte2vec           & 1-hot         & slice2vec         & 1-hot       & dep2vec          \\ \hline \hline
\textbf{64}                                                                    & 49.78\%            & 48.31\%            & 66.39\%         & 78.91\%          & 86.00\%      & 86.33\%        \\ \hline
\textbf{128}                                                                   & 53.01\%            & 53.52\%            & 68.51\%         & 80.57\%          & 84.81\%      & 87.75\%         \\ \hline
\textbf{256}                                                                   & 54.91\%            & 54.59\%            & 70.35\%         & 82.26\%          & 84.57\%      & 91.07\%           \\ \hline
\textbf{512}                                                                   & \textbf{55.80\%}            &    \textbf{55.96\%}                 &  \textbf{71.78}\% &           \textbf{83.35}\%        &  \textbf{86.34} \%      & \textbf{92.04\%}         \\ \hline \hline
\end{tabular}
\label{tab:task1}
\end{scriptsize}
\vspace{-1em}
\end{table}

\noindent
{\em Comparison setting.} There are two groups of comparison. First, we compare the in-set accuracy between the models trained on the bytecode sequences, program slices, and dependence paths to reveal the benefit of the program analysis. Second, we add the API elements embedding on the corresponding sequence corpora and compare them with the one-hot encoding counterparts. Altogether, we have three intermediate baselines, one for each of the bytecode sequences, slices, and dependence paths. This group aims to show the improvement of the embedding technique combined with program analysis. 

Tables~\ref{tab:task1} and ~\ref{tab:task2} evaluate the program-analysis guided embedding quality by the next API recommendation task and the next API sequence recommendation tasks, respectively. 
Compared with the basic one-hot on byte code (no program-specific designs are applied), the models trained with \textit{slice2vec} and \textit{dep2vec} achieve much higher accuracy in both tasks. The highest accuracies (92\% and 89\%) of both tasks are achieved by \textit{dep2vec}, which are 36\% and 46\% higher than their baseline settings, respectively. 

{\sloppy 
\noindent
\textit{Comparison between different program analysis techniques.}  Program analysis preprocessing techniques show significant benefits. For the next API recommendation task, the accuracy under one-hot encoding setting is improved from 56\% (one-hot on byte code) to 72\% (one-hot on slices) by inter-procedural slicing, and further to 86\% (one-hot on dependence paths) by the API dependence graph construction.  Under the embedding setting, the accuracy with \textit{slice2vec} is 27\% higher than \textit{byte2vec}.  The accuracy with \textit{dep2vec} is 92\%, 9\% higher than \textit{slice2vec}. The results of the next API sequence recommendation are also consistent with the conclusion. Comparison among the three one-hot settings shows that inter-procedural slicing improves the task accuracy by 20.5\%\footnote{64.10\% - 43.61\%=20.5\%}; API dependence graph construction improves the task accuracy by 18.8\%\footnote{82.94\% - 64.10\%=18.8\%}. \par}

\noindent
{\em Comparison between embedding and one-hot encoding.}  Results are shown in Table~\ref{tab:task1} and ~\ref{tab:task2}. There are significant improvements by applying embeddings on slices and dependence paths. In the next API recommendation task, \textit{slice2vec} improves the accuracy by 11\% from its one-hot baseline. \textit{dep2vec} improves the accuracy by 6\% from its one-hot baseline. These improvements suggest that \textit{slice2vec} and \textit{dep2vec} capture helpful information in vectors. In contrast, there is no significant improvement from applying embedding to the (unprocessed) byte code corpus. This conclusion is also observed in the next API sequence recommendation task. \textit{slice2vec} and \textit{dep2vec} improve the accuracy from their baselines by around 21\% and 6\%, respectively.  In contrast, \textit{byte2vec} does not show any significant improvement from its one-hot baseline.
 %

The impact of the LSTM units is as expected.  Longer LSTM units have higher accuracy.  Furthermore,  the accuracy increases more rapidly from LSTM-64 to LSTM-128 compared with LSTM-256 to LSTM-512.  It is likely because once the model has enough capacity, the differences caused by embedding are smaller.

 \begin{table}[t]
 \centering
 \caption{\small{Accuracy of the Next API Sequence Recommendation. We use the LSTM with hidden layer size 256 for this task.}}

\begin{tabular}{|b|a|b|a|b|a|}
\hline \hline
\rowcolor{Gray}
\multicolumn{2}{|c|}{\textbf{Byte Code}} & \multicolumn{2}{c|}{\textbf{Slices}} & \multicolumn{2}{c|}{\textbf{Dependence Paths}} \\ \hline 
1-hot           & byte2vec       & 1-hot        & slice2vec    & 1-hot    & dep2vec      \\ \hline \hline
43.61\%        & 44.63\%        & 64.10\%     & 85.02\%     & 82.94\%    & 89.23\%      \\ \hline \hline
\end{tabular}
\vspace{-1.5em}
\label{tab:task2}
\end{table}


\begin{figure}[t]
\centering
\includegraphics[width=\linewidth]{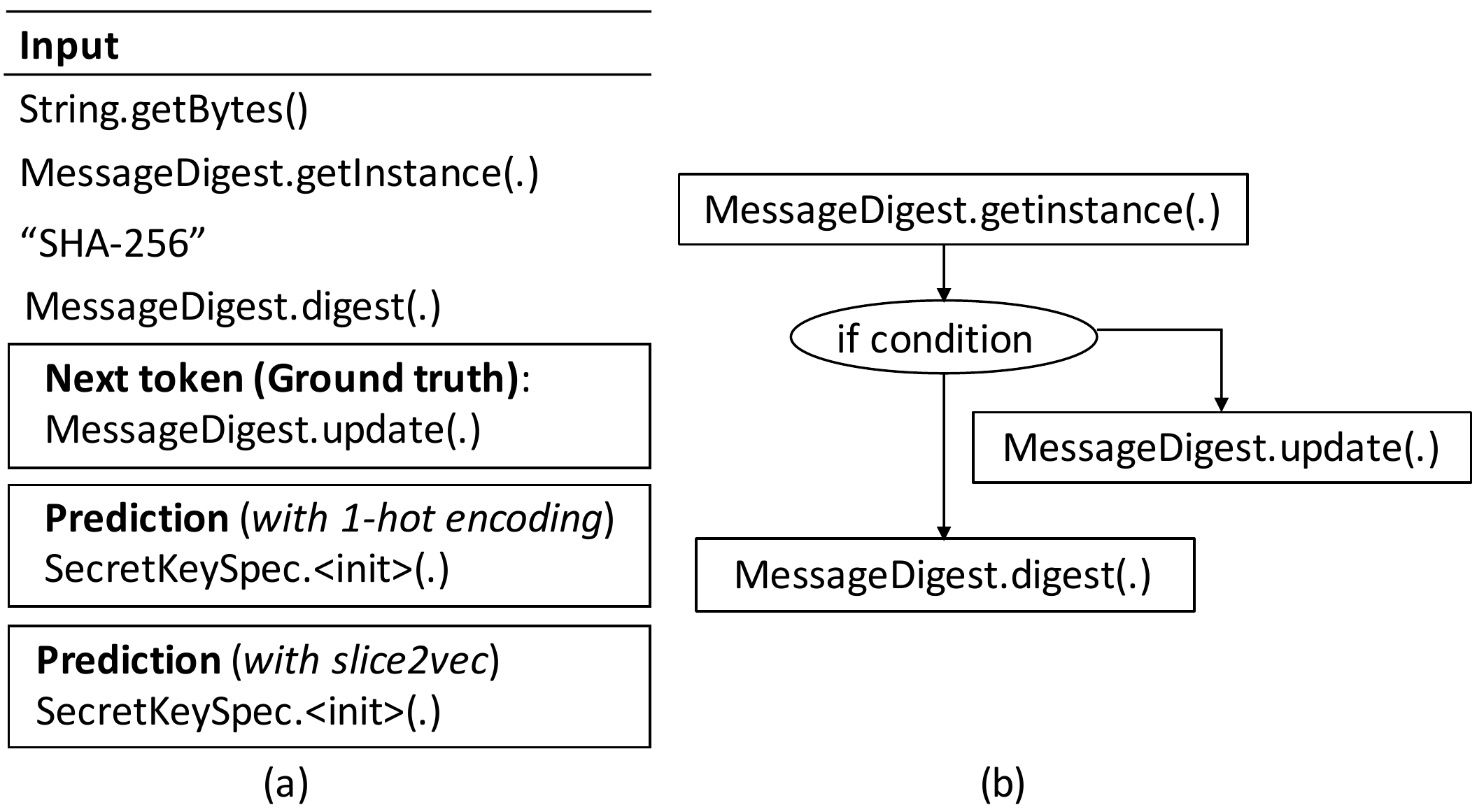}
\caption{\small{Case Study 1. (a) {\em slice2vec} and its one-hot slice baseline both incorrectly predict the test case. (b) {\em dep2vec} based on the dependency graph shown make the correct prediction.}}
\label{fig:case_study1}
\vspace{-1.5em}
\end{figure}

\smallskip
\noindent
{\sloppy
{\em Case Study 1.}  This case study is on the effectiveness of the API dependence graph construction. Figure~\ref{fig:case_study1}(a) shows a slice-based test case that is mispredicted by both {\em slice2vec} and its one-hot baseline. For digest calculation, it is common for \lstinline |MessageDigest.update(.)| to be followed by \lstinline |MessageDigest.digest(.)|, appearing 6,697 times in training. However, Figure~\ref{fig:case_study1}(a) shows a reverse order, which is caused by the \verb1if1-\verb1else1 branch shown in Figure~\ref{fig:case_study1}(b). When \lstinline |MessageDigest.update(.)| appears in an \verb1if1 branch, there is no guarantee which branch would appear first in slices. This reverse order is less frequent, appearing 1,720 times in training. Thanks to the API dependence graph construction, this confusion is eliminated, which predicts this case correctly. 


\par}

\begin{figure}[t]
    \centering

\includegraphics[width=\linewidth]{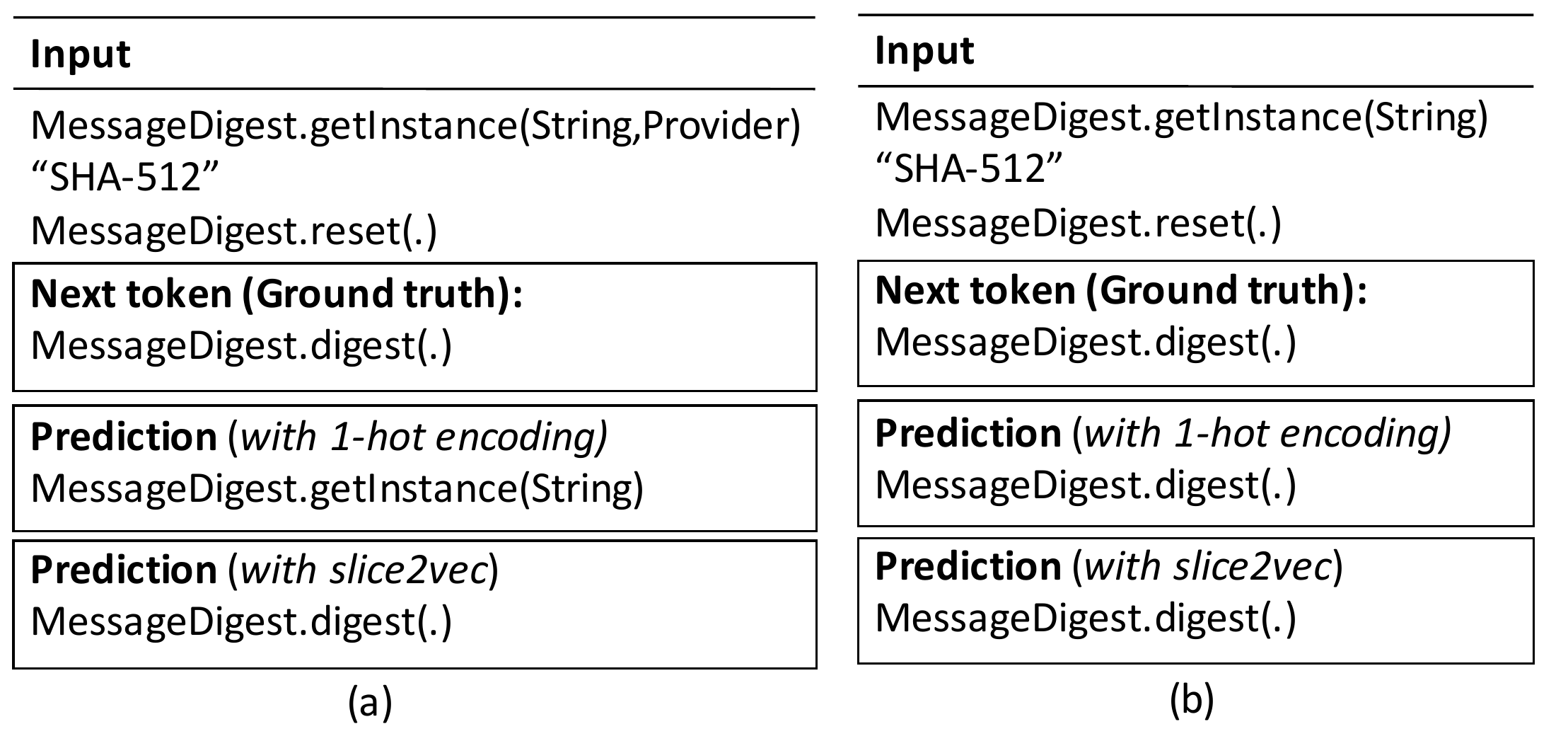}

\caption{\small{Case Study 2. (a) A test case that is correctly predicted by {\em slice2vec}, but incorrectly by one-hot encoding. (b) A test case similar to (a), but both {\em slice2vec} and one-hot give the correct prediction.}}
\label{fig:case_study2}
\vspace{-1.5em}
\end{figure}

\smallskip
\noindent
{\sloppy
{\em Case Study 2.} This case study is on the ability to recognize new previously unseen test cases. The slices in Figure~\ref{fig:case_study2}(a) and Figure~\ref{fig:case_study2}(b) slightly differ in the arguments of the first API. {\em slice2vec} makes the correct predictions in both cases, while its one-hot baseline fails in Figure~\ref{fig:case_study2}(a).
\lstinline |MessageDigest.getInstance(String)| appears much more frequent than \lstinline |MessageDigest.getInstance(String,Provider)| in our dataset. Specifically,  the former API appears 207,321 times, out of which 61,047 times are followed by the expected next token \lstinline |MessageDigest.digest(.)|. In contrast, the latter API -- where one-hot fails -- only appears 178 times, none of which is followed by \lstinline |MessageDigest.digest()|. 
In {\em slice2vec}, the cosine similarity between \lstinline |MessageDigest.getInstance(String,Provider)| and \lstinline |MessageDigest.getInstance(String)| is 0.68.~\footnote{For one-hot vectors, this similarity is 0.} This similarity, as the result of {\em slice2vec} embedding, substantially improves the model's ability to make inferences and recognize similar-yet-unseen cases. 
\par}

\begin{table}[b]
\centering
\vspace{-1em}
\caption{\small{The accuracy of HyLSTM for the next API recommendation. Acc.(A) refers to the in-set accuracy for all the test cases. Acc.(K) is the in-set accuracy for the known test cases. Acc.(U) is the in-set accuracy for new and unknown test cases.}}
\label{tab:hylstm}
\begin{small}
\vspace{-1em}
\begin{tabular}{|b|a|b|a|}
\hline \hline
\rowcolor{Gray}
                        & \textbf{HyLSTM}   & \begin{tabular}[c]{@{}c@{}}
                         \textbf{LSTM}      \\  (token-level loss)
                     
                        \end{tabular} & \begin{tabular}[c]{@{}c@{}}
                           \textbf{LSTM }   \\ (sequence-level loss)
                              
                        \end{tabular}\\ \hline \hline
Acc.(A)     & \textbf{93.00\%}  & 90.77\%    &  90.62\%     \\ \hline
Acc.(K)     & \textbf{99.86\%}      & 99.81\%     & 96.98\%       \\ \hline
Acc.(U)      & 56.94\%     &   43.21\%  &   \textbf{57.13\%}      \\ \hline \hline
\end{tabular}
\vspace{-1em}
\end{small}

\end{table}
\subsubsection{RQ3: How  to  recognize  low  frequency  long  API  sequences?}
We answer this research question with the comparative analysis within group 2 in Table~\ref{tab:comparative_setting}. 
We compare HyLSTM with two intermediate solutions, the LSTM with token-level loss and the LSTM with sequence-level loss, for the next API recommendation task.  The token-level loss is calculated based on the output at the last timestep while sequence-level combines the loss at every timestep of the sequence.
Their difference is that HyLSTM applies our low-frequency long-range enhancing technique while the others do not.  For all experiments, we use identical LSTM cells with hidden layer size 256.
We evaluate the {\it in-set accuracy} and further break it down as the in-set accuracy for known and unknown cases as defined below.


\smallskip
\noindent
{\em In-set accuracy of known cases.} Accuracy of known cases is the in-set accuracy calculated for the test cases that have appeared in the training.  We expect that the enhanced long-range influences could improve the accuracy of known cases.

\smallskip
\noindent
{\em In-set accuracy of unknown cases.} We calculate the in-set accuracy for new test cases unseen in the training set to see the inference capability of a model.  

As shown in Table~\ref{tab:hylstm}, HyLSTM outperforms the two intermediate solutions in the in-set accuracy by 1.23\%, and 1.38\%, respectively. The token-level loss and sequence-level loss show a trade-off between their accuracy for known cases and unknown cases. The HySLTM combines their strengths, substantially improving the accuracy for unknown cases by 13.73\%, compared with LSTM with token-level loss. HyLSTM also outperforms both intermediate solutions with the almost perfect accuracy for known cases at 99.86\%, without weakening their strengths. According to our manual case studies, most of the failed known cases are caused by the low-frequency API sequences. Therefore, higher accuracy for known cases without sacrificing the unknown cases gives a strong indication about the enhanced long-range influences.

\begin{figure}[t]
    \centering
    \includegraphics[width=.99\linewidth]{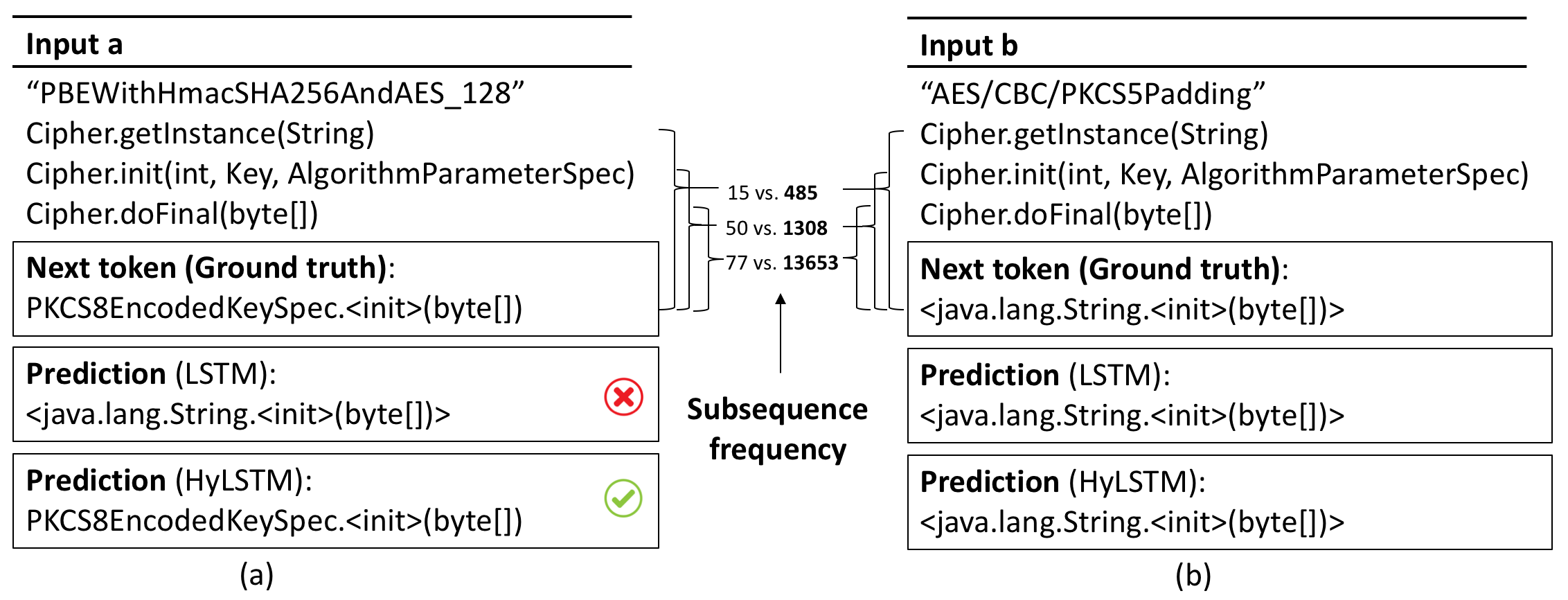}
    \caption{\small{Case Study 3, (a) A test case that is predicted incorrectly by regular LSTM and correctly by HyLSTM. (b) A test case that follows the frequent short pattern thus got correct prediction by both regular LSTM and our HyLSTM.}}
    \label{fig:case_study_global}
    \vspace{-1.5em}
\end{figure}

\noindent 
{\em Case Study 3.} This case verifies that our HyLSTM is better at identifying the low-frequency long-range patterns. Figure~\ref{fig:case_study_global} (a) shows a test case that is predicted incorrectly by regular LSTM but fixed by our HyLSTM. The regular LSTM makes the wrong prediction because the model is misled by the more frequent pattern shown in Figure~\ref{fig:case_study_global} (b). Compared with the label \lstinline|PKCS8EncodedKeySpec.<init>(byte[])|, the wrong prediction \lstinline|String.<init>(byte[])| is suggested more frequently according to these short suffixes. 
HyLSTM successfully differentiates the similar inputs from (a) and (b) with the long-range influences enhancing design. 


\subsubsection{RQ4: How  to  accurately  differentiate  different APIs that share the same functionality?}~\label{sec:multi-path}
We answer this question with the comparison within the group 3 in Table~\ref{tab:comparative_setting}. We compare our model with BERT, our multi-path architecture with their single-path counterparts in the next API recommendation task.
Table~\ref{tab:multi_paths} presents the in-set accuracy results for Multi-HyLSTM and the single-path counterparts. 
The Multi-HyLSTM achieves the best code in-set accuracy at 98.99\%, showing significant improvement (5.99\%) than the basic HyLSTM (Table~\ref{tab:hylstm}). 
\begin{table}[b]
\centering
\vspace{-1.5em}
\caption{\small{Comparison between multi-dependence suggestion and sequential suggestion. A, K, and U stand for in-set accuracy for all cases, known cases, and unknown cases, respectively.}}
\label{tab:multi_paths}
\begin{scriptsize}
\vspace{-1em}
\begin{tabular}{|l|a|b|a|b|}

\hline \hline
\rowcolor{Gray}
         &\textbf{Multi-HyLSTM} & \begin{tabular}[c]{@{}c@{}}\textbf{HyLSTM}\\ (path embedding)\end{tabular}  & \textbf{BERT}     & \textbf{Multi-BERT} \\ \hline \hline
Acc.(A)   & \textbf{98.99\%} & 95.79\% & 92.49\%     & 95.78\%     \\ \hline
Acc.(K)    &   99.59\%      & \textbf{99.84\%} & 99.48\%   & 96.52\%      \\ \hline
Acc.(U)     &   \textbf{83.02\%}     & 74.44\% & 55.73\%  & 76.07\%     \\ \hline \hline
\end{tabular}
\end{scriptsize}
\vspace{-1.5em}
\end{table}


\noindent
{\sloppy
{\em Multi-path vs. single-path}. Both Multi-HyLSTM and Multi-BERT are more accurate compared with their single-path counterparts. The in-set accuracy is improved from 95.79\% for HyLSTM with single path embedding to 98.99\% for Multi-HyLSTM, and from 92.49\% to 95.78\% for Multi-BERT. More importantly, multi-path aggregation gives significant accuracy improvement for previously unseen cases -- 9.58\% for Multi-HyLSTM and 20.24\% for Multi-BERT.
\par}


\noindent
{\em Improvement from path embedding}. The single-path embedding phase can benefit the API recommendation accuracy, especially for unknown cases. Compared with the basic HyLSTM in Table~\ref{tab:hylstm}, the extra path embedding phase improves the in-set accuracy from 93.00\% to 95.79\%, and from 56.94\% to 76.44\% for unknown cases.  This is likely because the extra path embedding phase exposes the model to the patterns that have not appeared in the task-specific training set.  

\noindent
{\em HyLSTM vs. BERT}. HyLSTM is better at learning programming languages compared with BERT. The HyLSTM outperforms BERT by 3.3\% in-set accuracy improvement and 18.71\% improvement for unknown cases.  The Multi-HyLSTM outperforms Multi-BERT by 3.21\% in-set accuracy improvement and 6.95\% improvement for unknown cases. 

\begin{figure}[t]
    \centering
    \includegraphics[width=.99\linewidth]{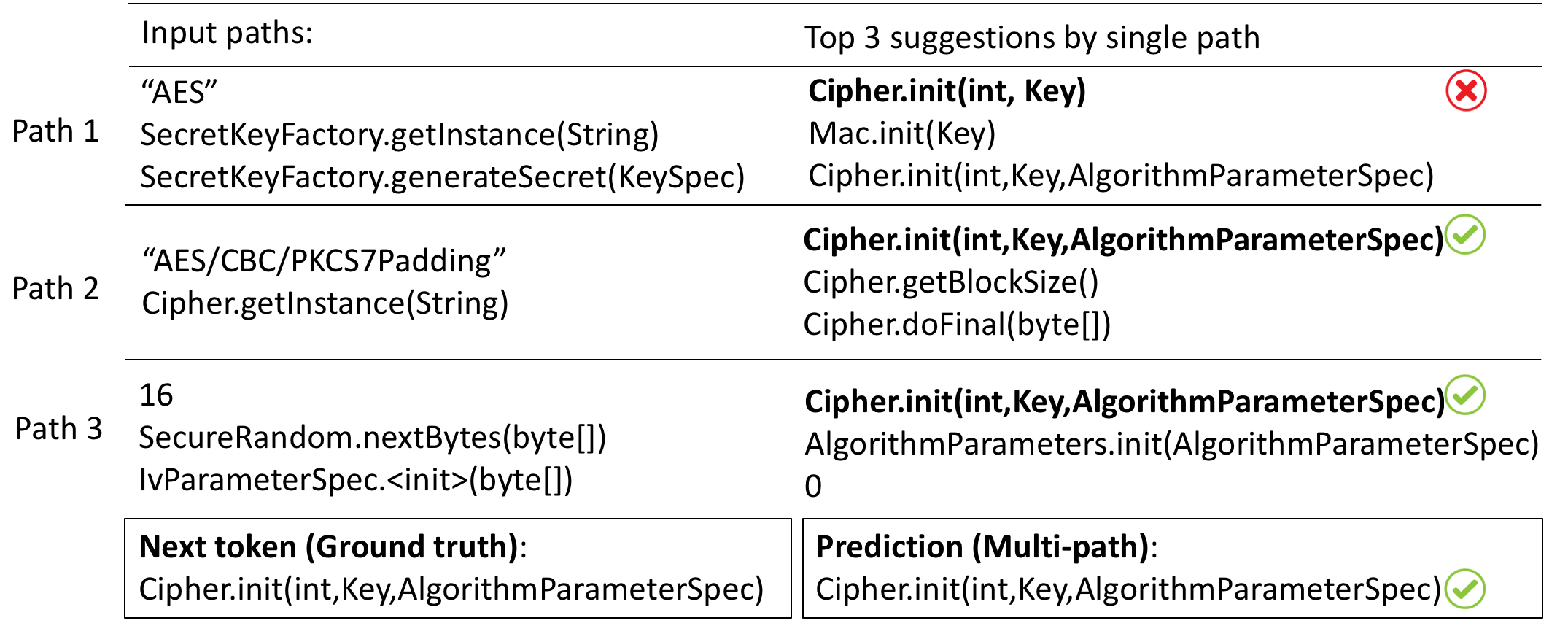}
    \vspace{-.5em}
    \caption{\small{Case Study 4. A test case that needs multiple paths to inference correct prediction. The wrong prediction suggested by Path 1 can be fixed after aggregating the influences from two extra paths.}}
    \label{fig:case_study_multi_path}
    \vspace{-2em}
\end{figure}

\noindent
{\em Case Study 4.} The case in Figure~\ref{fig:case_study_multi_path} demonstrates how the multi-path model improves over the single path model. The label \lstinline|Cipher.init(int,Key,AlgorithmParameterSpec)| and  \lstinline|Cipher.init(int,Key)| are indistinguishable, given dependence path 1. Fortunately, paths 2 and 3 provide complementary information to correct it.







\smallskip
\noindent
{\bf We summarize our experimental findings as follows:}

\begin{itemize}
   \item Our approach Multi-HyLSTM substantially outperforms the state-of-the-art API recommendation solutions SLANG and Codota in terms of the top-1 accuracy. Multi-HyLSTM achieves an excellent top-1 accuracy of 91.41\% compared with SLANG's 77.44\%. Compared with Codota that achieves a top-1 accuracy of 64.90\% in a manual analysis for 245 test cases, Multi-HyLSTM successfully recommends the 88.98\% cases with the top-1 answer.

   \item 
   
    Our multi-path architecture excels at recognizing new test cases that do not previously appear in the training set. Multi-HyLSTM and Multi-BERT improve the in-set accuracy for unknown cases  by 8.58\% and 20.34\% compared with their single-path counterparts, respectively.
    

   
   \item  HyLSTM with the low-frequency long-range enhancing design outperforms two regular LSTM models. 
   It improves the inference capability of the LSTM with token-level loss by 13.73\% and the memorizing capability of the LSTM with sequence-level loss by 2.88\%.  
   
    

    \item {\sloppy Program analysis guidance significantly improves the quality of code embedding and the recommendation accuracy. Enabled by these techniques, \textit{dep2vec} outperforms \textit{byte2vec} by 36\% for the next API recommendation, and 45\% for API sequence recommendation. \par}

    Similarly, the interprocedural slicing and API dependence graph construction also benefit in substantial accuracy improvements (31\% for the next API recommendation task and 39\% for the next API sequence recommendation task, respectively) even without embedding.

%
%

\end{itemize}

\smallskip
\noindent
{\em Performance and runtime.} With the distributed training of 8 workers, our training time is significantly improved. Most of our experiments are completed within 5 hours. We noticed that when the \textit{dep2vec} embedding is applied, the training becomes much faster than the one-hot encoding, improved from around 5 hours to 2 hours. 

\section{Threats to Validity and Limitations}

\noindent
{\bf Threats to Validity.}
A threat to internal validity is that the next token of a given sequence may not be the only choice. It may cause the prediction accuracy to be misaligned with the model capability. To avoid this issue, we collect a reasonable next token set to calculate accuracy. Although this set may not be complete, it reflects all the possibilities seen in the training phase. We think it is reasonable to evaluate the model according to the unseen knowledge rather than the full knowledge.    
Another threat to external validity is that our dataset is highly repetitive, which is different from other domains like images.  However, it is the characteristics of the code sequences. Different source code sequences may become identical after program analysis, which is expected. 

\noindent
{\bf Limitations.}
{\sloppy
\noindent
Our threat model described in Section~\ref{sec:challenges} can be further expanded to include other Java security libraries (e.g., Java Spring framework) and vulnerability types (e.g., TLS/SSL authentication-related vulnerabilities). We expect our pre-processing, embedding, and suggestion methodologies  to apply to these new scenarios.   
It is well known that static analysis tends to overestimate execution paths. Thus, the slices and dependence paths used for training embeddings and making suggestions might not necessarily occur. However, the infeasible paths appear together with the realizable paths. Our Multi-HyLSTM model can still make correct predictions, as it is decided by multiple data-flow paths. We expect the deep learning model to automatically learn which path to trust by training.
%
The dependence paths extracted from API dependence graphs may be incomplete, as we omit recursions in the call stacks and in the node paths. We also terminate the path when the depth of call stacks is beyond~10. A previous study experimentally showed the impact of limited depth exploration to be negligible in practice~\cite{rahaman2019cryptoguard}. 

%
%
\par}

\section{Other Related Work}


\noindent
\textit{Code completion work.} Many studies validated the naturalness of code sequences~\cite{hindle2012naturalness,tu2014localness,ray2016naturalness,hindle2016naturalness,allamanis2018survey,rahman2019natural}, which inspires a line of work to build language models for code completion/suggestion tasks~\cite{campbell2014syntax,bhatia2016automated,santos2018syntax,bhatia2018neuro,tufano2020generating,svyatkovskiy2020intellicode}.
%
%
For example, Campbell {\it et al.}~\cite{campbell2014syntax} utilized the $n$-gram model trained on the code tokens to localize and give suggestions for Java syntax error correction. Bhatia \textit{et al.}~\cite{bhatia2016automated,bhatia2018neuro} trained the Recurrent Neural Networks (RNN) on the student programming assignments to solve syntax errors in C++ by replacing or inserting suggested code tokens. Recently, a more advanced and powerful model, Transformer~\cite{vaswani2017attention}, is also applied to code generation tasks in ~\cite{svyatkovskiy2020intellicode,tufano2020generating}.  
These studies are built on top of the general source code token sequences or AST paths to solve syntax problems. However, as pointed in ~\cite{hellendoorn2017deep}, there do exist many program specific challenges that hinder the naive application of the language models borrowed from natural language processing domain. 

 Some studies focus on completion for API methods to improve the productivity of developers and solve API related problems~\cite{raychev2014code,nguyen2013statistical,nguyen2016api,nguyen2016learning}. Program analysis techniques are often applied to extract API sequences from source code to build language models. Nguyen \textit{et al.} presented a graph representation of object usage model (GORUM) to represent interactions between different objects and associated methods~\cite{nguyen2009graph}. They built Hidden Markov models for the state of objects and predict methods~\cite{nguyen2013statistical,nguyen2016learning}. However, these methods may require to build endless Markov models for different object types. Raychev \textit{et al.}~\cite{raychev2014code} built RNN and n-gram models on top of the object histories defined by themselves for API method recommendation. The object histories consist of the method call events in the temporal order. Although its top-16 accuracy (96.43\%) is pretty good, it only achieves a top-1 accuracy of 69.05\%.           

\noindent
{\em Code embedding work}.
Code embeddings (e.g.,~\cite{alon2019code2vec,henkel2018code,allamanis2015suggesting}) that automatically learn the vector representations of code tokens, have shown great improvement on the learning accuracy. There are multiple task-specific embedding studies (e.g., \cite{zhao2018deepsim,xu2017neural,alon2019code2vec,allamanis2015suggesting,li2018vuldeepecker}) which train the embedding jointly with the downstream task.  For example, Zhao {\it et al.}~\cite{zhao2018deepsim} trained embeddings of code snippets with the code similarity detection. 
Alon {\it et al.}~\cite{alon2019code2vec} designed a path-based attention network to train the code snippet embedding and predict its semantic tag (e.g., method name, class name). 
As these code vectors are optimized for specific tasks, they cannot be directly used in our API recommendation setting.







 Regardless of the downstream tasks, general-purpose embeddings are trained in the unsupervised way, e.g.,~\cite{nguyen2017exploring,harer2018automated}. The general-purpose embedding of source code in \cite{harer2018automated} improves the results of vulnerability detection compared with manual defined features. Recently, several embedding solutions for low-level code were  reported~\cite{ding2019asm2vec,ben2018neural}. Asm2Vec~\cite{ding2019asm2vec} embedded assembly functions based on their control flow graphs and apply embedding vectors for similarity detection between binaries. The authors in \cite{ben2018neural} learned embeddings for LLVM IR instructions and used it for algorithm classification and device mapping decisions (e.g., CPU vs. GPU). 
However, these designs for embedding source code tokens, low-level instructions, or LLVM IRs cannot be directly applied to API elements. 

\noindent
{\em Other API learning work.}
There is a line of work designed to solve API-related tasks~\cite{nguyen2016mapping,gu2016deep,nguyen2017exploring,chen2019mining,eberhardt2019unsupervised,bavishi2019autopandas,bui2019sar,xie2020api}. 
Nguyen {\it et al.}~\cite{nguyen2017exploring,nguyen2016mapping} mapped the semantic similar JDK APIs with .NET APIs through their similar embedding vectors. Chen {\it et al.}~\cite{chen2019mining} trained the API embedding based on the API description (name and documents)  and usage semantics to infer the likely analogical APIs between third party libraries. 
However, these solutions aimed at applying embeddings to help map analogical APIs, which is different from our API recommendation task with a high accuracy requirement. 
\section{Conclusions}

Data-driven code suggestion approaches need to be deeply integrated with program-specific techniques, as code and natural languages have fundamentally different characteristics.
We proposed new code suggestion techniques for securing Java cryptographic API usage, and comprehensively compared
our approach with a variety of state-of-the-arts and intermediate solutions. 
Our extensive evaluations with API recommendation and API sequence recommendation tasks demonstrated that our approach is more appropriate for programming languages than common sequential models designed for natural languages. 

\bibliographystyle{ACM-Reference-Format}
\bibliography{main}

\appendix
\section{Pseudo code}

We show the pseudo code for the entire training workflow of Multi-HyLSTM, and the multi-path selection algorithm below.
\begin{small}
\begin{algorithm}
	\caption{ModelTraining($D$): The workflow of training the embedding, HyLSTM and Multi-HyLSTM}
	\begin{algorithmic}[1]
	    \State{Input: $D$, where $D$ is a dataset of the API dependence graphs. Each graph $G$ has a slicing criterion $G.sc$}
	    \State{Output: $\theta$, where $\theta$ is the parameters of the trained Multi-HyLSTM}
	    \State $P$ $\leftarrow$ extract paths from $D$
	    \State $\theta_{emb} \leftarrow$ train $word2vec(P)$ //$\theta_{emb}$ is the parameters of the trained embedding layer.
	    \State $\theta_{HyLSTM} \leftarrow$ train $HyLSTM(P)$ //$\theta_{HyLSTM}$ is the parameters of the trained HyLSTM model
	     
	    \State $I \leftarrow \emptyset$ 
	    \State $L \leftarrow \emptyset$ 
	    \For {Graph $G$ in $D$}
	        \State $label \leftarrow G.sc$ 
	        \State $(P_1, \dots, P_n) \leftarrow$ MultiPathSection($G, label$)
	        \State $I.add((P_1, \dots, P_n))$
	        \State $L.add(label)$
	   \EndFor 
	    
	   \State //Initialize Multi-HyLSTM with $\theta_{emb}, \theta_{HyLSTM}$  
	   \State $\theta_{pretrain} \leftarrow (\theta_{emb}, \theta_{HyLSTM})$
	   \State //Start training
	   \For  {$(P_1, \dots, P_n)$ in $I$, $label$ in $L$}
	       \State $output \leftarrow$ Multi-HyLSTM$(P_1, \dots, P_n,\theta_{pretrain})$
	       \State $\theta \leftarrow$ backward\_propagation$(output,label) $
	   \EndFor
	\State \Return $\theta$
	\end{algorithmic} 
\end{algorithm} 
\begin{algorithm}
	\caption{MultiPathSelection($n$, $G$, $sc$): An important building block of ModelTraining for identifying n data-flow paths originating from the slicing criteria, with the constraint of being as non-overlapping as possible}
	\begin{algorithmic}[1]
	    \State{Input: ($n$, $G$, $sc$), where $n$ is the path budget and $G$ is an API dependence graph. $G$ includes a set of nodes $G.nodes$. A node $N$ contains a code statement $N.code$, the control dependence $N.cd$ of this code statement. $sc$ is a node in $G$ representing the slicing criterion. }
	    \State{Output: $C$, where $C$ includes $i$ data-flow paths ($i \le n$).}
	    \State  path $\leftarrow$ $\emptyset$
	    \State path.append($sc$)
		
		\State Q $\leftarrow$ $\emptyset$
		\State $Q$.enqueue($sc$)
		\While {$Q$ $\neq$ $\emptyset$}
		    \State $curr\_node \leftarrow Q$.dequeue()
		    \State $D_{control}$  $\leftarrow$ $\emptyset$
		    \For {Node $node$ in $curr\_node$.predecessor()}
		        \State $D_{control}$.add(node.cd)
		    \EndFor
		    \For {each $cd$ in $D_{control}$}
		        \For {Node $node$ in $curr\_node$.predecessor()}
		           \If {$node$.cd == $cd$} 
		               \State Q.enqueue($node$)
		               \State path.append($node$)
		               \State new\_path = PickAPath($path$, $G$)
		               \State C.add(new\_path)
		               \If {$C$.length == n}
		                   \State \Return
		              \EndIf
		           \EndIf
		      \EndFor
		    \EndFor
		\EndWhile

	\end{algorithmic} 
\end{algorithm} 

\begin{algorithm}
	\caption{PickAPath($path$, $G$): Traverse a data-flow path by random walk in an API dependence graph given several beginning nodes of the path}
	\begin{algorithmic}[1]
	    \State{Input: ($path$, $G$), where $path$ is an incomplete path only including several beginning nodes. $G$ is an API dependence graph.}
	    \State{Output: $new\_path$}
	    \State $new\_path \leftarrow \emptyset$ 
	    \State $curr\_node \leftarrow$ last node in $path$
	    \While {$curr\_node.predecessors() \neq \emptyset$ }
	        \State $next\_node \leftarrow$ randomly pick a predecessor 
	        \State $new\_path$.append($next\_node$)
	        \State $curr\_node \leftarrow next\_node$
	   \EndWhile
	   \State \Return $new\_path$
	\end{algorithmic} 
\end{algorithm} 

\end{small}

\end{document}